\documentclass[twocolumn,showpacs,aps,prl,superscriptaddress,floatfix]{revtex4}

\usepackage{graphicx}
\begin{document}
%
%Now come the authors with their affiliation
%
\author{S. R. Dunsiger}
\affiliation{Department of Physics and Astronomy, McMaster University, Hamilton, Ontario, L8S 4M1, Canada.}
\author{Y. Zhao}
\affiliation{Department of Physics and Astronomy, McMaster University, Hamilton, Ontario, L8S 4M1, Canada.}
\author{B. D. Gaulin}
\affiliation{Department of Physics and Astronomy, McMaster University, Hamilton, Ontario, L8S 4M1, Canada.}
\affiliation{Canadian Institute for Advanced Research, 180 Dundas St. W.,
Toronto, Ontario, M5G 1Z8, Canada}
\author{Y. Qiu}
\affiliation{National Institute of Standards and Technology, 
100 Bureau Drive, MS 6100, Gaithersburg, MD 20899-6102.}
\affiliation{Department of
Materials Science and Engineering, University of Maryland, College Park,
Maryland, 20742, USA}
\author{P. Bourges}
\affiliation{ Laboratoire Leon Brillouin, (UMR12 CEA-CNRS)
CEA-Saclay, 91191 Gif-sur-Yvette Cedex, France.}
\author{Y. Sidis}
\affiliation{ Laboratoire Leon Brillouin, (UMR12 CEA-CNRS)
CEA-Saclay, 91191 Gif-sur-Yvette Cedex, France.}
\author{J. R. D. Copley}
\affiliation{National Institute of Standards and Technology, 
100 Bureau Drive, MS 6100, Gaithersburg, MD 20899-6102.}
\author{A. B. Kallin}
\affiliation{Department of Physics and Astronomy, McMaster University, Hamilton, Ontario, L8S 4M1, Canada.}
\author{E. M. Mazurek}
\affiliation{Department of Physics and Astronomy, McMaster University, Hamilton, Ontario, L8S 4M1, Canada.}
\author{H.A. Dabkowska}
\affiliation{Department of Physics and Astronomy, McMaster University, Hamilton, Ontario, L8S 4M1, Canada.}
\title{Diagonal and Collinear Incommensurate Spin Structures in underdoped La$_{2-x}$Ba$_{x}$CuO$_{4}$}
\date{\today }
\newcommand{\LBCO}{La$_{2-x}$Ba$_x$CuO$_{4}$}
\newcommand{\LSCO}{La$_{2-x}$Sr$_x$CuO$_{4}$}
\newcommand{\LBCOy}{La$_{2-x}$Ba$_x$CuO$_{4+y}$}
\newcommand{\LSCOy}{La$_{2-x}$Sr$_x$CuO$_{4+y}$}
\newcommand{\tc}{T$_C$}
\begin{abstract}
We have studied incommensurate spin ordering in single crystal 
underdoped \LBCO\ with x$\sim$0.08, 0.05 and 0.025 using neutron scattering techniques.
Static incommensurate magnetic order is observed in the \LBCO\ (x=0.05 and 0.025) compounds with ordering wavevectors
which are rotated by 45$^{\circ }$ about the commensurate (0.5,0.5,0) %($\pi$, $\pi$) 
position, with respect to that in the 
superconducting x=0.08 material.
These spin modulations are one dimensional in the x=0.05 and 0.025 samples, with ordering wavevectors lying along the orthorhombic
$b^{*}$ direction.  
Such a rotation in the orientation of the static spin ordering as a function of increasing Ba doping, from diagonal to collinear, is roughly coincident with
the transition from an insulating to a superconducting ground state and is similar
to that observed in the 
related \LSCO\ system.  This phenomenon is therefore a generic property of underdoped La-214 cuprates.
\end{abstract}
\maketitle
%
%INTRO
% 
Lamellar copper oxides exhibit a wealth of fascinating phenomena which are a sensitive function of doping, 
evolving from an antiferromagnetic insulating N{\'e}el state through a Mott-Hubbard metal-insulator transition 
into a superconducting phase with 
increasing hole density~\cite{kastner98_tranquadareview_birgeneau06}. 
%A heterogeneous electronic phase composed of itinerant charges now appears to be a generic feature of the cuprates, 
Static and dynamic incommensurate spin states exist at all but the lowest doping levels in the cuprates, although
the true nature of the incommensurate spin ordered states is the subject of ongoing debate.
In an itinerant picture, the spin dynamics are described in terms of electron-hole pair
excitations about an underlying Fermi surface~\cite{bulut90_bulut96_norman00}.
Alternatively, within the ``stripe" picture of doped, two dimensional Mott insulators, the non-magnetic holes in these 
materials organize into quasi-one dimensional stripes which separate antiferromagnetic insulating antiphase 
domains~\cite{kivelson03}.  Adjacent antiferromagnetic regions are $\pi$ out of phase with each other 
giving rise to a magnetic structure with incommensurate periodicity, where the supercell dimension is twice the hole 
stripe periodicity.

The static spin structure in the undoped, parent compound
La$_2$CuO$_4$ has been determined by neutron scattering 
to be a simple two sublattice antiferromagnet
characterized by a commensurate ordering wavevector of (0.5, 0.5, 0)  
in reciprocal lattice units, or at the ($\pi$, $\pi$) position within the tetragonal basal plane~\cite{vaknin87}.
On hole doping with Sr$^{2+}$ substituting for La$^{3+}$ in La$_{2-x}$Sr$_x$CuO$_4$, 
the magnetic scattering moves out to incommensurate wavevectors~\cite{incommensurate1}.  
Lightly doped La$_{2-x}$Sr$_x$CuO$_4$ displays elastic incommensurate magnetic Bragg peaks which first 
appear split off
from the (0.5, 0.5, 0) position in diagonal directions relative to a tetragonal unit cell~\cite{matsuda00,wakimoto00}; that is at 
(0.5$\pm$$\delta \over \sqrt 2$, 0.5$\mp$$\delta \over \sqrt 2$, 0)
and (0.5$\pm$$\delta \over \sqrt 2$, 0.5$\pm$$\delta \over \sqrt 2$, 0) 
ordering wavevectors.
At higher doping in the underdoped superconducting regime, the Bragg peaks 
rotate by 45$^{\circ }$ to lie along directions parallel to or collinear with, the tetragonal axes or 
Cu-O-Cu bonds, such that
elastic magnetic scattering appears at (0.5$\pm$$\delta$,0.5,0) and (0.5,0.5$\pm$$\delta$,0)~\cite{suzuki98,kimura99}. 
For optimal and 
higher doping the static order disappears, but dynamic incommensurate correlations 
nevertheless persist\cite{kimura99,wakimoto04}.   

Surprisingly, La$_{2-x}$Ba$_x$CuO$_4$,  the first high temperature superconductor to be 
discovered~\cite{bednorz86}, has been 
much less extensively studied than \LSCO\ 
due to the difficulty of growing single crystals, which has only been recently achieved~\cite{adachi01}. 
 In this letter, we report the observation using neutron diffraction of  
the transition from a diagonal 
 to a collinear incommensurate spin ordering as an increasing function of doping in \LBCO. 
Elastic `collinear' incommensurate magnetic Bragg peaks are observed at T=1.5 K in superconducting \LBCO\ (x=0.08). 
In marked contrast, `diagonal' satellite peaks are observed at low temperature at reciprocal space positions rotated by 45$^{\circ}$ 
within the (HK0) plane for \LBCO\ (x$\sim $0.025 and 0.05).  This is analogous to the behaviour 
observed in the \LSCO\ 
materials and shows that such a rotation of the 
spin structure is a generic feature of the transition from insulating spin glass ground state to superconducting 
ground state in the underdoped La-214 cuprates.

\begin{figure}
\centering
\includegraphics[angle=90,width=1.0\columnwidth]{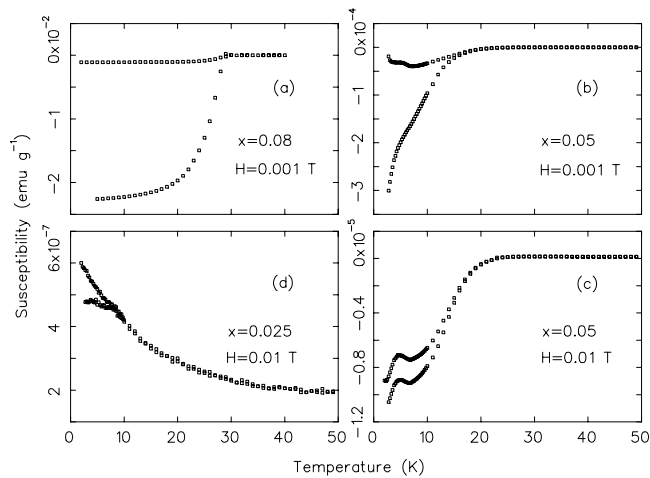}
\caption{Zero field cooled (ZFC) and field cooled (FC) susceptibilities of \LBCO\  a) x=0.08  b), c) x=0.05 and d) x=0.025 single crystals.} 
\end{figure}
We have grown high quality single crystals of \LBCO\ with  x=0.08, 0.05 and 0.025
using floating zone image furnace techniques with a four-mirror optical furnace~\cite{Yang, Hanna}.
The growth was performed in an oxygen atmosphere.
Samples of \LBCO\ near x=0.125 display a sequence of crystal structures on 
lowering the 
temperature, evolving progressively from high temperature tetragonal (HTT, I4/mmm) through orthorhombic (MTO, Bmab) to 
low temperature tetragonal (LTT, P4$_2$/ncm)~\cite{axe89}. 
The HTT to MTO structural phase transition at T$_{d1}$ is a sensitive indication of the precise Ba doping level, from which we 
have determined the Ba concentration in the x=0.08 sample (T$_{d1}$$\sim$305 K)~\cite{Yang}.  
It is more difficult to precisely quantify the doping levels in the x=0.05 and x=0.025 samples, where the crystals 
remain orthorhombic at low temperature.  As shown in Fig. 1, the diamagnetic superconducting 
volume fraction in the x=0.05 sample is roughly two orders of magnitude smaller than the x=0.08 sample, placing constraints on the doping 
level in the nominally x=0.05 sample~\cite{adachi01}.
Similarly, no signature of long range antiferromagnetic order is observed up to 300 K in the x=0.025 sample, setting a lower
limit on the doping level of x$>$0.02, by comparison with the suppression of the N{\'e}el order in \LSCO.

We explore both the spin glasslike and superconducting regimes of the phase diagram. 
At the highest Ba concentration, SQUID magnetisation measurements
indicate a bulk superconducting transition at T$_C$=29 K in the x=0.08 sample (see Fig 1a).
In \LBCO\ (x=0.05), a weak superconducting phase transition occurs 
around 20 K, as evidenced by the weak diamagnetic response (Fig 1b and 1c). Subsequently, history 
dependent behaviour develops below 10 K and a peak in the susceptibility is observed at 5 K, 
indicating a spin-glass ground state.
Finally, at the lowest Ba concentration, dc magnetization measurements  indicate the x=0.025 sample 
also undergoes a  spin glass-like transition below $\sim$10 K, as shown in Fig 1d).  

Time-of-flight neutron scattering measurements were
performed using the NG4 disk chopper spectrometer (DCS) at
the NIST Center
for Neutron Research. 
The DCS uses choppers to
create pulses of monochromatic neutrons whose energy
transfers on scattering are determined by their arrival 
times in the instrument's 913 detectors located at scattering
angles from -30 to 140$^{\circ }$. 
Measurements were performed
using 4.8 \AA\ incident neutrons for the x=0.08 and 0.05 samples,
allowing us to explore the area of reciprocal space around 
(0.5, 0.5, 0) with an energy resolution of 118 $\mu $eV FWHM.
Longer wavelength incident neutrons ($\lambda $=6.2 \AA\ with a corresponding energy resolution of 58 $\mu $eV FHWM) 
were used for the x=0.025 sample as the anticipated low 
incommensurability required higher Q resolution.
Definitive measurements of the incommensurate spin  structure are progressively 
more difficult at lower doping, as the incommensurability $\delta $ is proportional to the 
doping level x~\cite{yamada98}. 
%The time-of-flight technique is well suited to this problem, as it eliminates 
%any contamination from higher order neutrons which may otherwise feed through to the 
%(0.5, 0.5, 0) Bragg position.  
%Crystallographic indices 
%are denoted using
%tetragonal notation, where the basal plane lattice constant a=3.78 \AA\ at low temperatures. 
The measurements were carried out using an ILL orange cryostat with $^4$He 
exchange gas to ensure good thermal contact and each sample was aligned with the (H, K, 0) plane 
coincident with the horizontal scattering plane.

\begin{figure}
\centering
\includegraphics[angle=0,width=1.0\columnwidth]{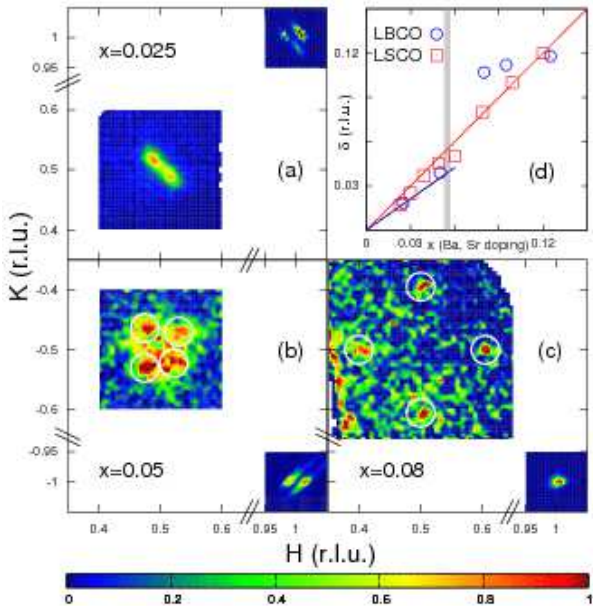}
\caption{Reciprocal lattice space maps in
the (H,K,0) plane, integrating over -0.1 meV $\le \hbar\omega \le $ 0.1 meV and taken at T$\sim $ 1.5 K in \LBCO\ x=0.025 (a),
0.05 (b) and 0.08 (c) respectively.  Diagonal magnetic Bragg scattering near (0.5$\pm$$\delta \over \sqrt 2$, 0.5$\mp$$\delta \over \sqrt 2$, 0) in the 
x=0.05 sample (b) and collinear magnetic Bragg scattering near (0.5$\pm$$\delta$,0.5,0) and (0.5,0.5$\pm$$\delta$,0) in the x=0.08 sample (c) have been circled 
for clarity. (d) Incommensurability, $\delta$ vs Ba/Sr concentration, x, plotted using the current results and those from the literature, as described in the text.  
The transition from diagonal to collinear incommensuration in the \LSCO\ system near x=0.055 \cite{matsuda00} is shown as the shaded vertical line.}
\end{figure}
Reciprocal space maps at T$\sim $ 1.5 K are shown in Fig 2, 
where in all cases we have integrated over the elastic scattering 
between 
-0.1 meV $\le \hbar\omega \le $ 0.1 meV.
%
%Fig. 2d) shows magnetic Bragg peaks in the \LBCO\ x=0.08 which occur at (0.5$\pm \delta $,0.5,0) and 
%(0.5, 0.5$\pm \delta$, 0) with  $\delta =0.107(3) $,
%indicating static {\it collinear} incommensurate spin order, with ordering wavevectors
%parallel to a$_{tetra}$. A single (1, 1, 0) structural Bragg peak is evident in 
%Fig 2 (d), indicating the x=0.08 sample is in its LTT phase at 1.5 K. 
%Complementary triple axis neutron measurements are described elsewhere~\cite{dunsiger07}.
%
Figure 2a) shows the elastic reciprocal space map from the lightly doped \LBCO\ (x=0.025) compound. These measurements were taken with
both 6.2 \AA\ and 8 \AA\ incident neutrons, in two different Brillouin zones, which gave consistent results 
(the 6.2 \AA\ data and the zone centred on (0.5, 0.5, 0) is shown here).
%yield a pattern of elastic 
%magnetic diffraction peaks very analogous to that observed in the x=0.05 sample.
A remarkable, isolated pair of {\it diagonal} incommensurate magnetic 
Bragg peaks are evident near (0.5$\mp$$\delta \over \sqrt 2$, 
0.5$\pm$$\delta 
\over \sqrt 2$, 0) with an incommensurability $\delta $=0.017(1), 
in tetragonal notation.  This is clear evidence for a static, one dimensional, diagonal incommensurate 
spin modulation along $b_{ortho}$.
Nuclear Bragg peaks associated with four MTO twin domains in the x=0.025 sample are visible near (1, 1, 0).  The integrated intensity of the majority twin peak
is $\sim$ 4 times larger than the minority peaks. This accounts for the single pair of incommensurate magnetic Bragg peaks: the 
magnetic scattering from this x=0.025 sample closely resembles that from a de-twinned MTO structure with a unique b-direction.

The same type of reciprocal space map is shown in Fig 2b) for the x=0.05 sample, now using $\lambda$=4.8 \AA\ incident neutrons.
Two nuclear Bragg peaks are now clear, centred around the tetragonal (1, -1, 0) position, associated with 
two out of four possible twin domains in the sample of \LBCO\ (x=0.05) at T=1.5 K in its MTO phase. 
The relative intensities of the two (1, -1, 0) peaks indicate the two twin domains have comparable volume fractions.
Magnetic Bragg peaks arising from static incommensurate spin order are again observed along the {\it diagonal} directions, that is
along each of the b$_{ortho}$ axes.  One pair of magnetic Bragg peaks is associated with each twin domain, such that a one dimensional spin
modulation occurs only along the orthorhombic b axis, as clearly illustrated in Figure 2b) of Ref~\onlinecite{wakimoto00}. 

%and the measured incommensurability, 
%$\delta$=0.389(6), is roughly twice that observed in the x=0.025 sample.

Such diagonal stripes have been predicted theoretically~\cite{machida89_kato90_poilblanc89_schulz89_zaanen90} and have also been 
observed in insulating
La$_{2-2x}$Sr$_x$NiO$_4$~\cite{tranquada96_yoshizawa00}.
The pattern is very similar to that observed in underdoped 
\LSCO\ (0.02$\le $x$\le $ 0.055)~\cite{wakimoto99,matsuda00,wakimoto00}, 
where the diagonal spin modulation is also along the orthorhombic
b axis and is considered
to be an intrinsic property of the entire insulating spin glass
region, in contrast to the parallel spin
modulation observed in the superconducting phase of \LSCO.
%Thus a drastic change takes place in
%the spin modulation vector, from diagonal to collinear, coincident with
%the lower critical concentration for superconductivity x$_c$
%$\approx $ 0.055~\cite{fujita02}.  
%These two types of
%spin modulations with similar incommensurabilities coexist near the boundary.

By contrast, magnetic Bragg peaks in \LBCO\ x=0.08 occur at (0.5$\pm \delta $,0.5,0) and 
(0.5, 0.5$\pm \delta$, 0) with  $\delta =0.107(3) $, as shown in the
map of reciprocal space in Fig 2c), again  measured using $\lambda$=4.8 
\AA\ incident neutrons. 
This indicates static {\it collinear} incommensurate spin order, with ordering wavevectors
parallel to a$^*_{tetra}$. A single (1, 1, 0) structural Bragg peak is evident in 
Fig 2 (c), indicating the x=0.08 sample is in its LTT phase at 1.5 K. 
Complementary thermal triple axis neutron measurements are described elsewhere~\cite{dunsiger07}.

The magnetic incommensurate ordering wavevectors, $\delta$ determined from the data in Figs 2(a)-(c) are 
plotted vs Ba concentration, x, 
in Fig. 2(d), along with the x-dependence of $\delta$ in \LSCO \cite{matsuda00}.  Data from
x=0.095 and 0.125 samples of \LBCO\cite{dunsiger07} are also included.  While minor differences are evident, 
the overall $\delta$ vs x behavior is very similar in this range 
of underdoped \LBCO\ and \LSCO.  The same transition, from diagonal to collinear incommensurate spin ordering is
roughly coincident with the transition from an insulating spin glass ground state to a superconducting ground 
state near x$_C$$\sim$ 0.055 for both \LBCO\ and \LSCO\ families, 
as shown by the vertical line in Fig. 2d).  The complex low temperature structural phase 
diagram of \LBCO\ clearly does not interfere 
with the rotation from diagonal to collinear incommensurate spin ordering as a function of doping.

\begin{figure}
\centering
\includegraphics[angle=90,width=1.0\columnwidth,clip]{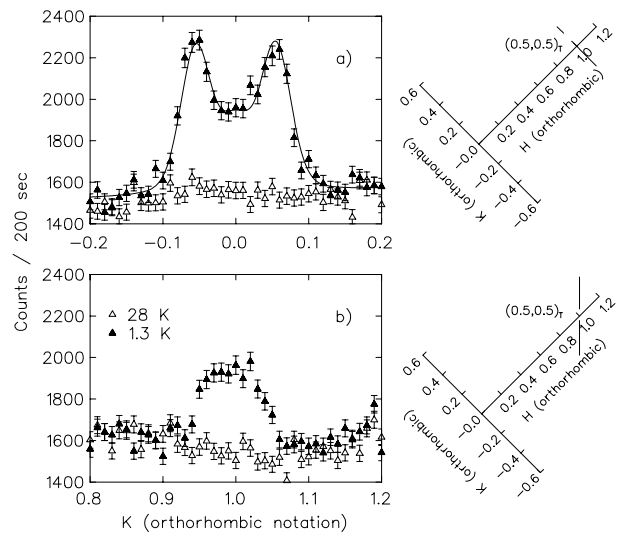}
\caption{Static incommensurate magnetic peaks with $\delta$=0.039 in La$_{1.95}$Ba$_{0.05}$CuO$_4$ at T=1.5 K, 
along a) the orthorhombic b$^{*}$ (1,K,0)$_O$ or `diagonal' direction, b) the `collinear' direction.  
The solid line in Fig 3a is discussed in the text.} 
\end{figure}

 Complementary cold triple axis neutron scattering experiments were undertaken on the same \LBCO\ (x=0.05) single 
crystal using 
the 4F1 spectrometer at the Laboratoire Leon Brillouin, France. The (002) reflection of pyrolytic graphite was used
for both monochromator and analyzer (E$_i$=E$_f$=5 meV). The latter was set in a flat configuration and no collimation was introduced,
 yielding an energy
resolution at $\hbar\omega =0$ of 105 $\mu$eV.  Cooled Be filters  were placed in both the incident and scattered 
beams to remove contamination from higher order neutrons.
The spectrometer was aligned to concentrate on a single twin domain. 
%with 
%lattice constants a=5.338 \AA\ and b=5.4125 \AA\
%(200) scans : peaks at 2.0001 and 1.9740
%(020) scans : peaks at 2.0013 and 2.0281
%or equivalently an orthorhombicity of $\frac{b}{a}-1=0.013957$.
Representative elastic scans along the orthorhombic and tetragonal b$^*$ axes are 
shown in Fig 3a) and b) respectively, using an orthorhombic coordinate system.
%The results of inelastic neutron scattering measurements on this concentration will be reported elsewhere.
Well resolved elastic magnetic peaks at an incommensurability $\delta $=0.0389(6) rlu in tetragonal notation ($\epsilon $=0.055(8) rlu$_{ortho}$) are 
clearly visible in Fig. 3a), corresponding to the static spin modulation along the b$^{*}$ direction.
By contrast, the intensity centred on the commensurate position in Figs. 3b) is due to the tails of the
incommensurate peaks coming from the b$^{*}$ modulation.
There is no evidence of any coexistence of diagonal and collinear spin ordering.

Quantitatively, the data in Fig 3a) have been analysed in terms of the 
%one dimensional 
convolution of the resolution with four Lorentzian functions
\begin{equation}
S(K)=\frac{A}{\pi } \sum _{n=1}^4
\frac{\kappa}{(K-K_{\delta, n})^2 +\kappa ^2} 
\end{equation}
where $K_{\delta, n}$ represents the four incommensurate wave vectors 
$(1,\pm \sqrt{2}\delta ,0)_O$, $(0.987 + \sqrt{2}\delta \cos\theta,0.0075 + \sqrt{2}\delta \sin\theta,0)_O$, and $(0.987 - 
\sqrt{2}\delta \cos\theta,0.0075 - \sqrt{2}\delta \sin\theta,0)_O$.
The two latter
contributions arise from the the wings of the second domain, characterised by a nuclear Bragg peak centred at $(1.974,0.015,0)_O$, 
such that $\tan\theta =0.015/1.974$.
The inverse of the static correlation length in the basal plane is defined as the peak half 
width at half maximum $\kappa $.
This functional form assumes the magnetic excitations consist of rods of scattering running along the c$^*$ axis.
The magnetic peaks are relatively sharp and practically resolution limited, with a HWHM of
0.0038(10) rlu in tetragonal notation, equivalent to a correlation length of 160(50) \AA\  within the basal plane.
This is of a comparable magnitude as that observed in the superconducting
state of \LBCO\ (x=0.095, 0.08)~\cite{dunsiger07}.  However, it should be noted that the values quoted 
for the related \LSCO\ x=0.05 compound are much shorter~\cite{wakimoto00,bao07}, 
between 25 - 35 \AA.

\begin{figure}
\centering
\includegraphics[angle=90,width=0.9\columnwidth,clip]{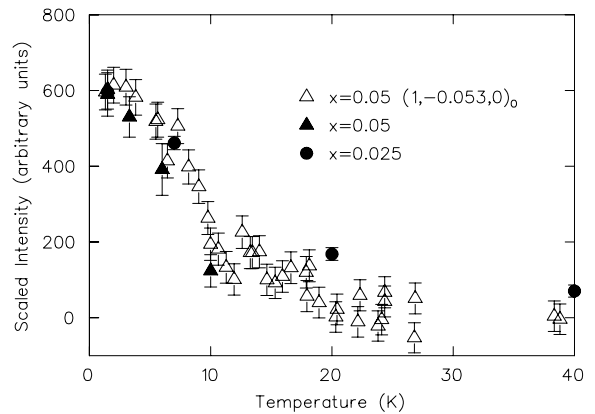}
\caption{
The temperature dependence of the net elastic incommensurate magnetic scattering in \LBCO\ (x=0.05) 
measured at (1,-0.053,0)$_O$ using triple axis (open triangles) and time-of-flight (filled triangles) 
neutron diffraction
techniques.  Integrated intensity between $-0.1 \le \hbar\omega \le 0.1$ meV  as a function of 
temperature in \LBCO\ (x=0.025)
measured using the DCS spectrometer (filled circles). 
}
\end{figure}
The temperature dependence of the incommensurate magnetic elastic scattering is 
shown in Fig. 4 for the x=0.05 and 0.025 samples, which display the diagonal incommensurate spin structures.   
Triple axis measurements 
of the magnetic Bragg intensity at (1,-0.053,0)$_O$ in the x=0.05 sample may be compared with DCS 
measurements of the intensity
integrated between $-0.1 \le \hbar\omega \le 0.1$ meV, as the energy resolution is similar.  
%Using neutrons of incident wavelength 4.8 \AA\ for the DCS
%measurements, the energy resolution is comparable (~0.1 meV FWHM) on the two instruments.
%
Figure 4 shows the decrease of the static incommensurate magnetic Bragg intensity in the x=0.05 sample on 
warming to T=10 K, consistent
with the temperature dependence of the break between the FC and ZFC susceptibilities shown in Fig. 1b).  
For temperatures beyond 10 K, a continued weak fall off of the intensity with increasing temperature is 
observed to $\sim$ 25 K, beyond which no
vestiges of the static signal are easily observable.  Our data for the temperature dependence of the 
incommensurate magnetic scattering in the
x=0.025 sample is less extensive, but qualitatively similar to that of the x=0.05 sample, as may be 
expected due to the similarity in the temperature
dependence of their ZFC vs FC susceptibilities and therefore their spin glass ground states.

% and indicates the magnetic
%ordering is static on the timescale of 4x10$^{-11}$ s below ~15 K, where history dependence develops in the magnetisation
%(see Fig 1b),c)), as anticipated in a spin glass-like system.
%
%

%

%Time-of-flight neutron scattering experiments on the lightly doped \LBCO\ (x=0.025) compound yield a 
%pattern of elastic 
%magnetic diffraction peaks very analogous to that observed in the x=0.05 sample.  As illustrated in Fig 2f), 
%nuclear Bragg
%peaks associated with four twin domains are visible.  However, note that the integrated intensity in the 
%majority peaks
% is between 3-5 times larger than the minority peaks.  Hence only a single pair of magnetic Bragg peaks
%are observed (Fig 2e), with an incommensurability $\delta $=0.017(1), of HWHM  0.033(4) rlu in tetragonal notation.
%This feature is associated with a static spin modulation along the orthorhombic b direction and 
%is seen at equivalent positions within two Brillouin zones using both 6.2 and 8 \AA\ incident 
%wavelength neutrons, lending weight that it is intrinsic to the material.
%As may be seen in Fig 4, the integrated intensity of the magnetic Bragg peaks drops gradually as 
%the temperature is increased, though it remains finite above the spin glass transition at 10 K.

%
In conclusion, we have observed static, {\it diagonal}, one dimensional incommensurate spin ordering  
in La$_{2-x}$Ba$_x$CuO$_4$ with x=0.05 and x=0.025, evolving into {\it collinear} incommensurate order in 
the superconducting x=0.08 sample.  This result is very similar to the evolution of incommensurate spin order 
in the well studied underdoped system, La$_{2-x}$Sr$_{x}$CuO$_4$, implying the phenomenon is a generic 
feature of the underdoped 
La-214 cuprates.  

It is a pleasure to acknowledge the contributions of A. Dabkowski.  This work
utilized facilities supported in part by the National Science Foundation
under Agreement No. DMR-0454672.  We acknowledge the support of NSERC of Canada. 


\begin{thebibliography}{41}
%
\bibitem{kastner98_tranquadareview_birgeneau06} M. A. Kastner {\it et al}, Rev Mod Phys {\bf 70}, 897 (1998);  J. M. Tranquada, cond-mat/0512115 (2005); R.J. Birgeneau {\it el al}, J Phys Soc Jpn (2006).
%
%\bibitem{kastner98} M. A. Kastner {\it et al}, Rev Mod Phys {\bf 70}, 897 (1998).
%
%\bibitem{tranquadareview} J. M. Tranquada, cond-mat/0512115 (2005).
%
%\bibitem{birgeneau06} R.J. Birgeneau, C.Stock, J.M. Tranquada , K.Yamada, J Phys Soc Jpn (2006).
%
\bibitem{bulut90_bulut96_norman00} N. Bulut, D. Hone, D.J. Scalapino, 
N.E. Bickers, Phys Rev Lett 
{\bf 64}, 2723 (1990); N. Bulut, D. J. Scalapino, Phys Rev B {\bf 53}, 5149 (1996); M. R. Norman, Phys Rev B {\bf 61}, 14751 (2000).
%\bibitem{bulut90} N. Bulut, D. Hone, D. J. Scalapino,N. E. Bickers, Phys Rev Lett {\bf 64}, 2723 (1990). 
%
%\bibitem{bulut96} N. Bulut, D. J. Scalapino, Phys Rev B {\bf 53}, 5149 (1996).
%
%\bibitem{norman00} M. R. Norman, Phys Rev B {\bf 61}, 14751 (2000).
%
\bibitem{kivelson03} S. A. Kivelson {\it et al}, Rev Mod Phys {\bf 75}, 1201 (2003).
%
\bibitem{vaknin87} D. Vaknin {\it et al}, Phys Rev Lett {\bf 58}, 2802 (1987).
%
\bibitem{incommensurate1} K. Yamada {\it et al}, Phys. Rev. B {\bf 57}, 6165 (1998); H. Kimura {\it et al}, 
Phys. Rev. B {\bf 59}, 6517 (1999). 
%
%\bibitem{matsuda00_wakimoto00} M. Matsuda {\it et al}, Phys. Rev. B {\bf 62}, 9148 (2000); 
%S. Wakimoto {\it et al}, Phys. Rev. B {\bf 61}, 3699 (2000).
%
\bibitem{matsuda00} M. Matsuda {\it et al}, Phys. Rev. B {\bf 62}, 9148 (2000); 
%
\bibitem{wakimoto00} S. Wakimoto {\it et al}, Phys. Rev. B {\bf 61}, 3699 (2000).
%
\bibitem{suzuki98} T. Suzuki {\it et al}, Phys Rev B {\bf 57}, R3229 
(1998).  
%
%\bibitem{suzuki98} T. Suzuki, T. Goto, K. Chiba, T. Shinoda, T. Fukase, H. Kimura, K. Yamada, M. Ohashi, Y. Yamaguchi, Phys Rev B {\bf 57}, 3229 (1998).  
%
\bibitem{kimura99} H. Kimura {\it et al}, Phys Rev B {\bf 59}, 6517 (1999).
%
%\bibitem{kimura99} H. Kimura, K. Hirota, H. Matsushita, K. Yamada, Y. Endoh, S-H Lee, C. F. Majkrzak, R. Erwin, G. Shirane, M. Greven, Y. S. Lee, M. A. Kastner and R. J. Birgeneau, Phys Rev B {\bf 59}, 6517 (1999).%
\bibitem{wakimoto04} S. Wakimoto {\it et al}, Phys Rev Lett, {\bf 92} 217004 (2004).
%
%\bibitem{tranquada95} J. M. Tranquada, B. J. Sternlieb, J. D. Axe, Y. Nakamura, S. Uchida, Nature {\bf 375}, 561 (1995).
%
%\bibitem{tranquada97} J. M. Tranquada, J. D. Axe, N. Ichikawa, A. R. Moodenbaugh, Y. Nakamura, S. Uchida, Phys Rev lett {\bf 78}, 338 (1997). 
%
%\bibitem{fujita02} M. Fujita, H. Goka, K. Yamada, M. Matsuda,Phys Rev B, {\bf 66} 184503 (2002).
%
\bibitem{bednorz86} J.G. Bednorz and K.A. Muller, Z. Phys B {\bf 64}, 189 (1986).
%
\bibitem{adachi01} T. Adachi, T. Noji, and Y. Koike, Phys Rev B {\bf 64}, 
144524 (2001).
%
%\bibitem{fujita04} M. Fujita, H. Goka, K. Yamada, J. M. Tranquada, L. P. Regnault, Phys Rev B {\bf 70}, 104517 (2004).
%
%\bibitem{moodenbaugh88} A. R. Moodenbaugh {\it et al}, Phys Rev B {\bf 38}, 4596 (1988).
%
\bibitem{Yang} Y. Zhao {\it et al}, Phys Rev B {\bf 76}, 184121 (2007).
%
\bibitem{Hanna} H.A. Dabkowksa and B.D. Gaulin, J. Optoelec. Adv. Mat., {\bf 9}, 1215 (2007).
%
\bibitem{axe89} J. D. Axe  {\it et al}, Phys Rev Lett {\bf 62}, 2751 (1989).
%
\bibitem{yamada98} K. Yamada {\it et al}, Phys Rev B {\bf 57}, 6165 (1998). 
%
%\bibitem{yamada98} K. Yamada, C. H. Lee, K. Kurahashi, J. Wada, S. Wakimoto, S. Ueki, H. Kimura, Y. Endoh,S. Hosoya, G. Shirane, R. J. Birgeneau, M. Greven, M. A. Kastner, and Y. J. Kim, Phys Rev B {\bf 57}, 6165 (1998). 
%
%\bibitem{matsuda02} M. Matsuda, M. Fujita, K. Yamada, R. J. Birgeneau, Y. Endoh, G. Shirane, Phys Rev B {\bf 66}, 174508 (2002).
%
%\bibitem{yamada89} K. Yamada, K. Kakurai, Y. Endoh, T. R. Thurston, M. A. Kastner, R. J. Birgeneau, G. Shirane, Y. Hidaka, T. Murakami, Phys Rev B {\bf 40}, 4557 (1989).
%
%\bibitem{shirane87} G. Shirane, Y. Endoh, R. J. Birgeneau, M. A. Kastner, Y. Hidaka, M. Oda, M. Suzuki, and T. Murakami, Phys. Rev. Lett. {\bf 59}, 1613 (1987).
%
%\bibitem{yamada95} K. Yamada, S. Wakimoto, G. Shirane, C. H. Lee, M. A. Kastner, S. Hosoya, M. Greven, Y. Endoh, R. J. Birgeneau, Phys Rev Lett {\bf 75}, 1626 (1995).
%
%\bibitem{yamada97} K. Yamada, C. H. Lee, Y. endoh, G. Shirane, R. J. Birgeneau, M. A. Kastner, Physica C {\bf 282-287}, 85 (1997)
%
%\bibitem{hiraka01} Haruhiro Hiraka, Yasuo Endoh, Masaki Fujita, Young S. Lee, Jiri Kulda, Alexandre Ivanov, Robert J. Birgeneau, J. Phys Soc Jpn {\bf 70}, 853 (2001).
%
%\bibitem{fujita02a} M. Fujita, K. Yamada, H. Hiraka, P. M. Gehring, S. H. Lee, S. Wakimoto and G. Shirane,  Phys Rev B {\bf 65}, 064505 (2002).
%
%\bibitem{savici02} A. T. Savici, Y. Fudamoto, I. M. Gat, T. Ito, M. I. Larkin, Y. J. Uemura, G. M. Luke, K. M. Kojima, Y. S. Lee, M. A. Kastner, R. J. Birgeneau, K. Yamada, phys Rev B {\bf 66}, 014524 (2002).%
%
\bibitem{machida89_kato90_poilblanc89_schulz89_zaanen90} K. Machida, Physica C {\bf 158}, 192 (1989); 
M. Kato {\it et al}, J. Phys. Soc. Jpn. {\bf 59}, 1047 (1990); 
D. Poilblanc and T.M. Rice, Phys. Rev. B {\bf 39}, 9749 (1989); 
H. Schulz, J. Phys. France {\bf 50}, 2833 (1989); 
J. Zaanen and O. Gunnarsson, Phys. Rev. B {\bf 40}, 7391 (1989).
%\bibitem{machida89} K. Machida, Physica C {\bf 158}, 192 (1989).
%
%\bibitem{kato90} M. Kato, K. Machida, H. Nakanishi, and M. Fujita, J. Phys. Soc. Jpn. {\bf 59}, 1047 (1990).
%
%\bibitem{poilblanc89} D. Poilblanc and T.M. Rice, Phys. Rev. B {\bf 39}, 9749 (1989).
%
%\bibitem{schulz89} H. Schulz, J. Phys. France {\bf 50}, 2833 (1989).
%
%\bibitem{zaanen90} J. Zaanen and O. Gunnarsson, Phys. Rev. B {\bf 40}, 7391 (1990).
%
\bibitem{tranquada96_yoshizawa00} J.M. Tranquada, D.J. Buttrey, and V. Sachan, Phys. Rev. B {\bf 54},
12 318 (1996); H. Yoshizawa {\it et al}, Phys. Rev. B {\bf 61}, R854 (2000).
%
%\bibitem{tranquada96} J.M. Tranquada, D.J. Buttrey, and V. Sachan, Phys. Rev. B {\bf 54}, 12 318 (1996).
%
%\bibitem{yoshizawa00} H. Yoshizawa, T. Kakeshita, R. Kajimoto, T. Tanabe, T. Katsufuji, and Y. Tokura, Phys. Rev. B {\bf 61}, R854 (2000).
%
\bibitem{wakimoto99} S. Wakimoto {\it et al}, Phys. Rev. B {\bf 60}, R769 (1999).
%
%\bibitem{wakimoto99} S. Wakimoto, G. Shirane, Y. Endoh, K. Hirota, S. Ueki, K. Yamada, R.J. Birgeneau, M.A. Kastner, Y.S. Lee, P.M. Gehring, and S.H. Lee, Phys. Rev. B {\bf 60}, R769 (1999).
%
\bibitem{dunsiger07} S. R. Dunsiger {\it et al}, arXiv:0708.2519.
%
\bibitem{bao07} W Bao {\it et al}, Phys. Rev. B {\bf 76} 180406(R) (2007).
%
\end{thebibliography}
\end{document}